\documentclass[%
 reprint,
 amsmath,amssymb,
 aps,
 prl,
]{revtex4-2}

\makeatletter
\let\@bibdataout@init\relax
\AtBeginDocument{\let\pre@bibdata\@empty}
\makeatother

\usepackage{graphicx}
\usepackage{bm}
\usepackage{xcolor}
\usepackage{tikz}
\usetikzlibrary{arrows.meta,positioning,fit,backgrounds}
\usepackage{hyperref}
\hypersetup{colorlinks=true,linkcolor={blue!55!black},
  citecolor={blue!55!black},urlcolor={teal!70!black}}

\usepackage{listings}
\lstdefinestyle{lean}{
  basicstyle=\ttfamily\normalsize,
  columns=fullflexible,
  keepspaces=true,
  showstringspaces=false,
  literate=
    {ℕ}{{$\mathbb{N}$}}1
    {ℝ}{{$\mathbb{R}$}}1
    {∣}{{$\mid$}}1
    {≤}{{$\leq$}}1
    {∃}{{$\exists$}}1
    {→}{{$\rightarrow$}}1
    {γ}{{$\gamma$}}1
    {β}{{$\beta$}}1
}

\usepackage{amsthm}
\newtheorem{conjecture}{Conjecture}

\newcommand{\eres}{\varepsilon^{\mathrm{res}}}
\newcommand{\Z}{\sigma^z}
\newcommand{\X}{\sigma^x}
\newcommand{\bhat}{\hat{\bm b}}
\newcommand{\zhat}{\hat{\bm z}}
\newcommand{\tauv}{\bm\tau}
\newcommand{\mvec}{\bm m}
\newcommand{\SU}{\mathrm{SU}(2)}
\newcommand{\SO}{\mathrm{SO}(3)}

\definecolor{leanblue}{HTML}{1A85FF}
\newcommand{\leancheck}{\tikz[scale=0.22, line cap=round, line join=round, baseline=0pt]{
  \draw[leanblue, line width=0.9pt, shift={(-0.45,0.05)}] (0,0.3) -- (0.28,0) -- (0.95,0.75);
  \draw[white, line width=2.1pt] (0,0.3) -- (0.28,0) -- (0.95,0.75);
  \draw[leanblue, line width=0.9pt] (0,0.3) -- (0.28,0) -- (0.95,0.75);}}

\begin{document}

\title{A Machine-Verified Proof of a Quantum-Optimization Conjecture}

\author{Uri Kol}
\affiliation{Center of Mathematical Sciences and Applications,
 Harvard University, Cambridge, Massachusetts 02138, USA}

\author{Maor Ben-Shahar}
\affiliation{MIT Center for Theoretical Physics - a Leinweber Institute, Cambridge, MA 02139, USA}

\author{Kfir Sulimany}
\thanks{Corresponding author}
\email{Kfir@mit.edu}
\affiliation{Research Laboratory of Electronics, Massachusetts Institute of Technology, Cambridge, Massachusetts 02139, USA}

\author{Dirk Englund}
\affiliation{Research Laboratory of Electronics, Massachusetts Institute of Technology, Cambridge, Massachusetts 02139, USA}

\date{\today}

\begin{abstract}
We report a machine-verified resolution of a problem open for over a decade in quantum optimization: the Farhi, Goldstone and Gutmann (FGG) conjecture \cite{Farhi:2014ych} that depth-$p$ Quantum Approximate Optimization Algorithm (QAOA) on the ring of disagrees attains approximation ratio $(2p+1)/(2p+2)$ exactly. We found the proof using a large language model, Claude Fable~5, and verified its correctness end-to-end by the Lean~4 proof assistant. Our methodology includes several ingredients: building on a substantial Lean library of quantum information, we formalized the QAOA components and the known parts of the problem, and reduced the conjecture to a single open mathematical statement.
The model was then handed the library and our agentic toolkit, and tasked with closing that gap by constructing a proof in Lean.
The resulting process is a feedback loop between the model's natural-language reasoning and Lean's mechanical verification, which converged to a machine-verified proof.
Human verification is required only for the structural scaffolding - that the formal statement faithfully encodes the intended claim - while the proof itself is supplied by the model and certified mechanically by Lean. The proof is nevertheless striking - the model uncovered a hidden dynamical symmetry of the problem and exploited it, borrowing tools and machinery from an adjacent field to turn a hard existence problem into an explicit construction. This work paves the way for resolving open conjectures in quantum information science and beyond.
\end{abstract}

\maketitle

\section{Introduction}\label{sec:intro}
Large language models are beginning to contribute directly to scientific reasoning: they can propose calculations, identify patterns, generate code, and search for proofs. Their speed makes it possible to explore arguments that would be costly to examine one by one. For mathematically precise problems the generation of an argument can be separated from its certification. A model may propose a proof, while an independent program checks whether every step follows from explicitly stated assumptions.

Such verification is essential because large language models can hallucinate: they may invent facts or lemmas, overlook exceptional cases, or introduce subtly invalid inferences while producing fluent and persuasive explanations. The danger is especially acute in long mathematical arguments, where one incorrect step can invalidate everything that follows and may be difficult to identify by inspection. Readability, confidence, and even agreement among several models are no substitutes for deterministic verification.

Formal proof systems such as Lean~4 \cite{deMouraUllrich2021,mathlib2020} can be used as a tool for verification of AI proposed theorems. With such formal systems a proof that is accepted (namely, compiles cleanly) establishes that its conclusion follows from its hypotheses and any of the axioms that are assumed. What Lean does not establish is that the formal statement expresses the intended claim. The semantic gap, however, is confined to the comparatively compact statement of the theorem and its definitions, which is the only place where human verification is required. The proof connecting hypotheses to conclusion is delegated in full to the machine.

This generate-then-certify paradigm has recently produced a wave of machine-generated results across mathematics: olympiad geometry~\cite{Trinh2024alphageometry}, silver-medal olympiad problem solving in Lean~\cite{AlphaProof2025}, and algorithmic discovery~\cite{RomeraParedes2024funsearch}. These build on advances in autoformalization~\cite{Wu2022autoformalization} and language-model theorem proving~\cite{Yang2023leandojo,DeepSeekProverV2,DraftSketchProve}, including agentic Lean provers that extend to the physical sciences~\cite{AxProver2025}.

We apply this paradigm to a long-standing open problem in quantum optimization, which dates to the original analysis of the Quantum Approximate Optimization Algorithm (QAOA) by Farhi, Goldstone, and Gutmann (FGG). They identified a particularly clean benchmark known as the \emph{ring of disagrees} \cite{Farhi:2014ych}. In this problem, \(n\) binary spins are placed around a ring, and the objective is to maximize the number of neighboring pairs that point in opposite directions. In optimization language, this is the MaxCut problem on a cycle; in physics language, it is a one-dimensional antiferromagnetic Ising model with periodic boundary conditions. When \(n\) is even, the optimum is the alternating spin configuration, in which every neighboring pair disagrees. A depth-\(p\) QAOA circuit prepares a quantum state by alternating \(p\) evolutions generated by the Ising cost Hamiltonian with \(p\) evolutions generated by a transverse-field mixing Hamiltonian.
The approximation ratio \(r_p\) is the cost expectation value in the QAOA state, normalized by the true optimum. The optimal approximation ratio \(r_p^\ast\) is the largest expected fraction obtainable by varying the \(2p\) evolution angles. FGG conjectured that, for even \(n\) in the regime~\mbox{\(2p+2\leq n\)},
\begin{equation}
  r_p^\ast=\frac{2p+1}{2p+2}.
  \label{eq:qaoa-ring-ratio}
\end{equation}
Settling this conjecture would pin the exact expressive power of depth-\(p\) QAOA on this model: what the quantum circuit can achieve when its parameters are chosen optimally, namely an \emph{optimal performance guarantee}. Such guarantees are the canonical figure of merit for optimization algorithms, and in particular for QAOA on MaxCut~\cite{Wurtz2021guarantees}. Beyond its theoretical interest, QAOA and related combinatorial-optimization protocols are now executed on programmable quantum hardware, including superconducting processors~\cite{Harrigan2021} and neutral-atom Rydberg arrays~\cite{Pichler2018,Ebadi2022,Nguyen2023}; the one-dimensional antiferromagnetic Ising ring studied here is itself natively realized on such arrays~\cite{Bernien2017,Keesling2019}.

In this work, we combine human-guided autoformalization, custom Lean infrastructure, and the large language model Claude Fable~5 to obtain a machine-verified proof of the FGG conjecture~\eqref{eq:qaoa-ring-ratio}. We formalized the optimization problem, constructed the definitions and supporting library required by the prover, and isolated the open gap as a single formal statement. Claude Fable~5 generated the decisive argument, which Lean~4 accepts with the standard classical axioms used in Mathlib \cite{mathlib2020}. The result is therefore an independently certified proof, not one resting on trust in either the model-generated output or an informal derivation. The complete Lean development is publicly available~\cite{repo}, and for completeness we also present the proof itself below.

\section{QAOA and the ring of disagrees}\label{sec:review}

\subsection{The algorithm and the benchmark}

The QAOA \cite{Farhi:2014ych} is a hybrid quantum-classical ansatz for combinatorial optimization. A classical cost function is encoded in a Hamiltonian $H_C$ which, together with a mixing operator $B$, is used to construct the  depth-$p$ QAOA state
\begin{equation}
  |\bm\gamma,\bm\beta\rangle
  = e^{-i\beta_p B}e^{-i\gamma_p H_C}\cdots e^{-i\beta_1 B}e^{-i\gamma_1 H_C}\,
    |+\rangle^{\otimes n},
\end{equation}
with the $2p$ angles $(\bm\gamma,\bm\beta)$ chosen to minimize $\langle H_C\rangle$. The standard terminology in the combinatorial-optimization literature uses a cost operator, related to the Hamiltonian, whose spectrum is $[E_{\min},E_{\max}]$, by $C=(E_{\max}-H_C)/2$, so $C$ is maximal at the ground state, where $C_{\max}=\tfrac12(E_{\max}-E_{\min})$. The approximation ratio compares the achieved cost to this optimum,
\begin{equation}
  r_p = \frac{\langle C\rangle}{C_{\max}}
      = \frac{E_{\max}-\langle H_C\rangle}{E_{\max}-E_{\min}} =: 1-\eres_p,
\end{equation}
so $r_p=1$ at the ground state, and the normalized \emph{residual energy} $\eres_p$ is the shortfall. The object of interest is the optimum over angles
\begin{equation}
r_p^\ast=\max_{\bm\gamma,\bm\beta} r_p(\bm\gamma,\bm\beta).
\end{equation}

The benchmark we focus on is the \emph{ring of disagrees}, MaxCut on the cycle graph, with cost and mixer
\begin{equation}
  H_C = \sum_{j=1}^{n}\Z_j\,\Z_{j+1}\quad,
  \qquad B = \sum_{j}\X_j .
\end{equation}
Its spectrum is $[-n,n]$, and $C=(n-H_C)/2$ counts disagreeing edges. We take $n$ even, so the antiferromagnetic ground state is unfrustrated and $C_{\max}=n$.

\subsection{Exact decomposition and the open gap}

FGG predicted $r_p^\ast$ on this benchmark from numerics and proved the case $p=1$. Wang \emph{et al.} \cite{Wang:2018rtf} extended the proof to $p=2$ by an exact fermionic analysis. Mbeng, Fazio, and Santoro \cite{Mbeng:2019rat} derived an exact decomposition formula of the residual energy for any \(p\) in terms of independent momentum modes, which we summarize here since the open gap in the proof lies there.

The ring is effectively a free-fermion model~\cite{LiebSchultzMattis1961}. A Jordan-Wigner transformation~\cite{JordanWigner1928} maps it to non-interacting spinless fermions, and translational invariance block-diagonalizes the dynamics into $p$ decoupled \emph{momentum modes} at
\begin{equation}\label{eq:momenta}
  k_\ell = \frac{\ell\pi}{p+1}, \qquad \ell = 1,\dots,p.
\end{equation}
On each mode the \emph{Anderson pseudospin}~\cite{Anderson1958} reduces the $(k,-k)$ fermion pair to a single spin-$\tfrac12$ with Pauli operators $\tauv_k = (\tau^x_k,\tau^y_k,\tau^z_k)$. In this representation, the mixer and cost operators act, mode by mode, as two competing Zeeman fields,
\begin{equation}\label{eq:per-mode-ham}
\begin{gathered}
  H_C^{(k)} = -2\,\bhat_k\cdot\tauv_k, \qquad B^{(k)} = 2\,\zhat\cdot\tauv_k,\\
  \bhat_k = (-\sin k,\,0,\,\cos k),
\end{gathered}
\end{equation}
the mixer along $\zhat$ and the cost along the \emph{cost axis} $\bhat_k$. Since the modes are independent, the residual energy splits into a sum over them \cite{Mbeng:2019rat},
\begin{equation}\label{eq:decomp}
  \eres_p = \frac{1}{2p+2} + \frac{1}{2p+2}\sum_k \varepsilon_k, \qquad \varepsilon_k\ge 0,
\end{equation}
and each per-mode residual is a single-qubit geometric quantity,
\begin{equation}\label{eq:eps-geom}
  \varepsilon_k = 1 - \bhat_k\cdot\mvec_k = \tfrac12\,\lVert \mvec_k - \bhat_k\rVert^2,
\end{equation}
where $\mvec_k=\langle\tauv_k\rangle$ is the pseudospin's \emph{magnetization}. In the QAOA state $\mvec_k$ is its initial value $\zhat$, acted on by the $p$ layers, an alternating product of $\SO$ rotations about $\zhat$ (mixer) and $\bhat_k$ (cost),
\begin{equation}\label{eq:mvec}
  \mvec_k = \Big(\prod_{\ell=p}^{1} R_{\zhat}(4\beta_\ell)\,R_{\bhat_k}(4\gamma_\ell)\Big)\zhat .
\end{equation}
$\varepsilon_k$ is therefore the misalignment between $\mvec_k$ and its cost axis $\bhat_k$, vanishing exactly when the layers steer the mode onto that axis, $\mvec_k=\bhat_k$.

Each $\varepsilon_k\ge 0$, so \eqref{eq:decomp} gives $\eres_p\ge 1/(2p+2)$ \cite{Mbeng:2019rou}, that is $r_p^\ast\le (2p+1)/(2p+2)$. FGG conjectured that this bound is saturated.

\begin{conjecture}\label{conj:main}
\emph{(FGG \cite{Farhi:2014ych}).} For the ring of disagrees with $n$ even and
$2p+2\le n$, the optimal depth-$p$ QAOA approximation ratio is exactly
$r_p^\ast = (2p+1)/(2p+2)$.
\end{conjecture}

By \eqref{eq:decomp} the bound is saturated exactly when there exist angles that simultaneously drive every mode onto its cost axis,
\begin{equation}\label{eq:gap}
  \exists\,\bm\gamma,\bm\beta:\ \forall k,\ \varepsilon_k(\bm\gamma,\bm\beta)=0 ,
\end{equation}
and this simultaneous-steering problem is the gap in the literature. Each condition $\varepsilon_k=0$ pins a unit Bloch vector to a target axis - that is two real constraints, so the $p$ modes give $2p$ equations in the $2p$ angles - a square system. Equal counts would force a solution if the equations were linear, but they are transcendental in the angles through the $\SO$ rotations of \eqref{eq:mvec}, and a square transcendental system may have no real solution, one, or many. Mbeng, Fazio, and Santoro confirmed saturation numerically to $p=128$ and gave an explicit schedule only for $2p\ge n$; analytic variational protocols for transverse-field Ising chains were given by Ho and Hsieh~\cite{HoHsieh2019}. The complementary regime $2p+2\le n$, where the $p$ momenta \eqref{eq:momenta} are the relevant modes \cite{Farhi2020needs}, is the one their schedule leaves open and the regime of the conjecture; there they inferred existence from the parameter count, which is not a proof \cite{Mbeng:2019rat,Bravyi:2019jtd}. Supplying that existence proof is what closes the gap.

\section{Machine-proved result}\label{sec:approach}

\begin{figure*}[t]
\centering
\hspace*{-1mm}
\begin{tikzpicture}[
  >={Triangle[length=2.5mm, width=1.9mm]}, font=\footnotesize,
  node distance=16mm and 18mm,
  box/.style={rounded corners=2pt, draw, align=center, inner sep=4pt,
              minimum height=14mm, minimum width=23mm, text width=23mm},
  llm/.style={box, fill=blue!7},
  comp/.style={box, fill=black!8},
  good/.style={box, fill=green!14},
  inp/.style={box, fill=cyan!20, minimum width=15mm, text width=15mm},
  aux/.style={box, fill=orange!14},
  lab/.style={font=\footnotesize, align=center, inner sep=2pt},
  act/.style={font=\footnotesize\itshape},
  e/.style={->, thick}
]
\exhyphenpenalty=50 \hyphenpenalty=50 
\node[llm]                      (plan) {Natural-language\\[2pt] proof plan};
\node[inp,  left=8mm of plan]  (input) {Input\\[2pt] problem};
\node[aux,  below=10mm of plan] (py)   {Python\\[2pt] unit-testing};
\node[llm,  right=19mm of plan] (code) {Lean~4\\[2pt] formalization};
\node[comp, right=16mm of code] (comp) {Lean~4 compiler\\[2pt] \& kernel};
\node[good, right=16mm of comp] (cert) {Machine-checked\\[2pt] proof};

\draw[{Triangle[length=1.8mm, width=1.4mm]}-{Triangle[length=1.8mm, width=1.4mm]}, thick, shorten >=0.5mm, shorten <=0.5mm] (plan) -- node[lab, left=-10.0mm, font=\scriptsize]{numerical\hspace{0.9em}checks} (py);
\draw[e, shorten >=0.5mm] (plan) -- node[lab, left=0.5mm, above, align=center, yshift=3pt]{formalize}(code);

\draw[e, shorten >=0.5mm] (input) -- (plan);

\draw[e, shorten >=0.5mm] (code) -- node[lab, left=1mm, above, yshift=1pt]{compile} (comp);
\draw[e, shorten >=0.5mm] (comp) -- node[lab, left=1mm, above, yshift=1pt]{accept} (cert);

\begin{scope}[on background layer]
  \node[draw, dashed, rounded corners, fill=blue!3, inner sep=7pt,
        fit=(plan)(code)(py)] (llmbox) {};
  \draw[white, line width=1.6pt]
        ([yshift=-1mm]llmbox.east |- code) -- ([yshift=4mm]llmbox.east |- code);
  \draw[white, line width=2.4pt]
        ([yshift=-2.2mm]llmbox.west |- plan) -- ([yshift=1.5mm]llmbox.west |- plan);
\end{scope}
\node[act, anchor=south east] at ([shift={(-2mm,1.5mm)}]llmbox.south east) {LLM (Claude Fable~5)};
\draw[white, line width=1.6pt]
      ([xshift=-2.2mm]llmbox.north -| plan.north) -- ([xshift=2.2mm]llmbox.north -| plan.north);
\draw[e] (comp.north) to[out=90, in=90, looseness=0.7]
      node[lab, above, yshift=+3pt]{compiler diagnostics:}
      node[lab, below, yshift=-3pt]{type errors \& open goals} ([yshift=0.5mm]plan.north);
\node[act, below=2mm of comp.south] {Verifier (Lean~4)};
\node[anchor=south east, inner sep=0.8mm] at (cert.south east) {\leancheck};
\end{tikzpicture}
\caption{\textbf{The autoformalization feedback loop.} The input is a formalized statement, here the open conjecture on the optimal performance of depth-$p$ QAOA on the Ising ring, handed to the system to prove. The LLM (Claude Fable~5 in our case) drafts a proof strategy in natural language, validating its candidate statements with lightweight Python numerical checks as part of the planning process.
The Python checks are internal to the model and never reach Lean.
Only a plan that passes these internal checks is formalized as Lean~4 code.
The Lean compiler and kernel - the external verifier - then check the attempt and return diagnostics, namely type errors and the remaining proof goals, which the model uses to revise the plan. The cycle repeats until Lean accepts the proof, yielding a terminal machine-checked certificate.}
\label{fig:feedback}
\end{figure*}
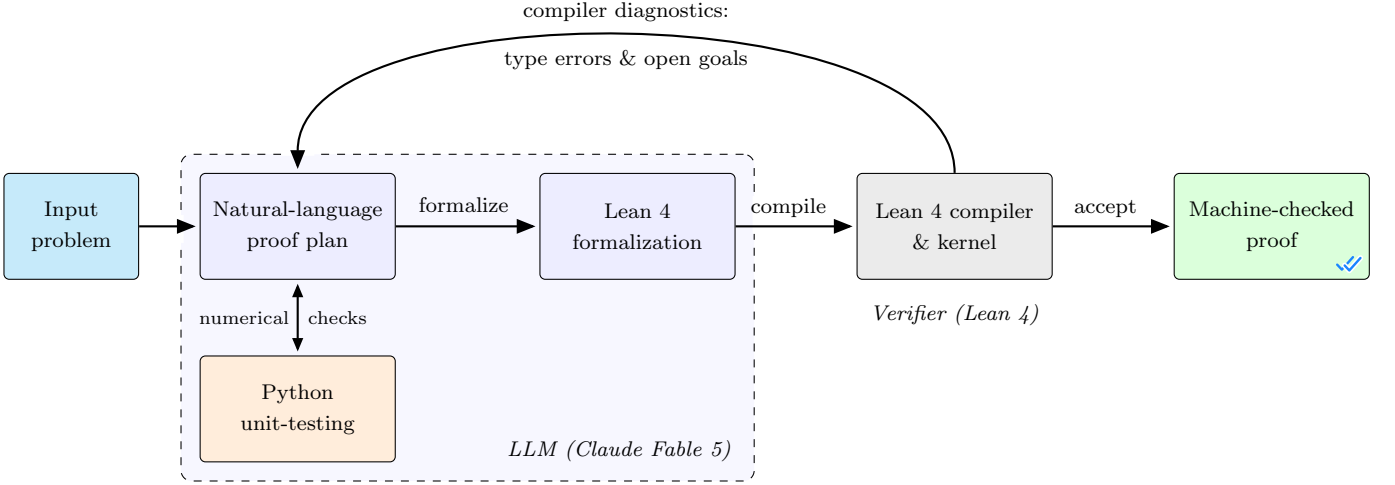

To supply that proof, we built on a Lean formalization of quantum information that we developed in our parallel work on machine-verified quantum cryptography~\cite{benshahar_qcrypto} - which provided the quantum-mechanical definitions and supporting mathematical theorems that underlie the present formalization - together with parts of the agentic autoformalization toolkit~\cite{leanscriber}.

In order for the model to be able to state the conjectured theorem, we extended this library with the further terms and definitions it required. The inherited formalization already supplied the basic quantum mechanical objects - bras, kets, density operators, tensor products, and inner products - so our additions were the machinery specific to QAOA, such as the mixing Hamiltonians and the depth-$p$ state ansatz.
We have also formalized the known components of the problem - the definition of the Ising chain, including its exact decomposition into decoupled momentum modes and the supporting machinery (Jordan-Wigner transformation and Anderson's pseudospin) - using various tools, including our agentic toolkit for autoformalization \cite{leanscriber}.

The gap \eqref{eq:gap} was then given to Claude Fable~5, with the same toolkit \cite{leanscriber}, and it was tasked with closing the theorem by constructing a proof in Lean. The LLM made natural language proof plans in the process and attempted to formalize them, thus obtaining compiler feedback on its proof attempts. This workflow is described in Fig.~\ref{fig:feedback}: the loop takes a formalized open statement as its input and carries it through to a machine-checked proof. Many such statements can be posed; here it is the existence statement \eqref{eq:gap} of the FGG conjecture. Although Lean compilation was the final verification tool for the LLM, it also used python for numerical checks before investing proof efforts in any conjectured construction, in order to make sure that its theorem statements were not obviously false.

The conjecture is encoded by the Lean theorem \texttt{residualEnergy\_isLeast} \cite{repo}, which states that $1/(2p+2)$ is the \emph{least} attainable residual energy - equivalently, that the optimal approximation ratio equals $(2p+1)/(2p+2)$:
\begin{lstlisting}[style=lean]
theorem residualEnergy_isLeast {N P : ℕ}
    (hN_even : 2 ∣ N) (hP : 2 * P + 2 ≤ N) :
    IsLeast {e | ∃ γ β : Fin P → ℝ,
        residualEnergy (ringQAOA N) γ β = e}
      (1 / (2 * P + 2))
\end{lstlisting}
Its proof combines two theorems:
the \emph{lower bound} - $\eres_p\ge 1/(2p+2)$ for all angles (\texttt{residualEnergy\_lower\_bound}) - and \emph{attainability} - the explicit angle construction attaining the optimal value (\texttt{residualEnergy\_attained}).
Both are fully formalized and checked by the kernel rather than assumed. We have checked that these three theorems, together with \texttt{residualEnergy}, \texttt{ringQAOA}, and all the definitions they depend on, faithfully encode the conjecture. The statement and its definitions were fixed before the model saw the gap, and the model never changed them. The proof was filled by Claude Fable~5 and it compiles cleanly with only the classical axiom set used in Mathlib.

\section{The proof}\label{sec:proof}

With a faithful formalization of the theorem and a machine certification provided by Lean, the validity of the proof is independent of whether any human understands it. Nevertheless, the proof is striking in its own right, and we describe it below.

By the decomposition \eqref{eq:decomp} the problem is whether some choice of the $2p$ angles makes every per-mode residual vanish. The proof rests on a bridge to another problem: each momentum mode's propagator is a \emph{Quantum Signal Processing} (QSP) sequence, which turns the steering condition into a condition on a single polynomial.

\subsection{A symmetry principle}\label{sec:symmetry}

The proof relies on a symmetry of the per-mode dynamics that is not manifest in the original formulation of the problem.

\paragraph{The $\SO$ symmetry.}
By the per-mode reduction, the $p$ layers act not on the full $2^n$-dimensional state but on a single Bloch vector $\mvec_k$ per mode, rotated by the alternating product of $\SO$ elements \eqref{eq:mvec}. The reachable per-mode dynamics is that of the rotation group on a three-vector, rather than the generic, exponentially large QAOA evolution.

\paragraph{Lifting to $\SU$.}
Rotations of a Bloch vector are the adjoint action of the double cover $\SU$: with $\tauv_k$ as the spin-$\tfrac12$ generators, each $R_{\hat n}(\theta)\in\SO$ is the image of $e^{-i(\theta/2)\hat n\cdot\tauv_k}\in\SU$ under~\mbox{$\mathrm{Ad}(U)(\hat v\cdot\tauv_k)=U(\hat v\cdot\tauv_k)U^\dagger=(R\hat v)\cdot\tauv_k$}. Two $\SU$ elements carry the per-mode evolution - the depth-$p$ propagator $U_k$ and a fixed frame rotation $W_k$,
\begin{equation}\label{eq:Uk}
  U_k = \prod_{\ell=p}^{1} e^{-2i\beta_\ell\,\tau^z_k}\,e^{-2i\gamma_\ell\,\bhat_k\cdot\tauv_k},
  \qquad W_k = e^{-ik\,\tau^y_k/2},
\end{equation}
whose adjoint actions on $\zhat$ furnish the magnetization and the cost axis,
\begin{equation}\label{eq:m-adjoint}
  \mvec_k = \mathrm{Ad}(U_k)\,\zhat , \qquad \bhat_k = \mathrm{Ad}(W_k^\dagger)\,\zhat .
\end{equation}
With both now $\SU$ objects, this lift is the entry point into QSP - the theory of $\SU$-valued sequences built from alternating generators.

\paragraph{Dynamical symmetry.}
$\SU$ is not a symmetry of the Hamiltonian but a \emph{dynamical} symmetry. The cost Hamiltonian singles out $\bhat_k$, so $\SO$ does not commute with $H_C$. What holds instead is that the mixer and cost generators lie in a closed three-dimensional algebra, $\mathrm{span}\{\tau^x_k,\tau^y_k,\tau^z_k\}\cong\mathfrak{su}(2)$, confining the dynamics to $\SU$. That the dynamical Lie algebra (DLA) is this small is exceptional: for a generic QAOA instance the DLA fills a substantial part of $\mathfrak{su}(2^n)$, and the evolution explores exponentially many directions. Here it collapses because the model is free - Jordan-Wigner maps $H_C$ and $B$ to quadratic operators, whose bilinears close into $\mathfrak{so}(2n)$, and translational invariance diagonalizes that algebra into $\mathfrak{su}(2)$ blocks. The entire DLA is thus $\bigoplus_k\mathfrak{su}(2)_k$, of dimension $O(p)$ rather than $O(4^n)$. The DLA viewpoint on variational circuits is due to Larocca \emph{et al.} \cite{Larocca:2021ksf} and, for free-fermionic circuits, Matos \emph{et al.} \cite{Matos:2022koi}; the dynamical Lie algebras of translation-invariant free-fermion chains have since been fully classified~\cite{Wiersema2024dla}. Because the dynamics preserve this block structure, the problem reduces to independent single-qubit rotations, which is what makes a single-qubit theory like QSP the right tool.

\subsection{The keystone}\label{sec:keystone}

Consider the dot product $\bhat_k\cdot\mvec_k$ of the cost axis and the magnetization, both $\SU$ adjoint images of $\zhat$. Since the adjoint action preserves dot products, defining the
\emph{lifted propagator}
\begin{equation}\label{eq:Gdef}
  G(k) := W_k\,U_k \in \SU
\end{equation}
gives $\bhat_k\cdot\mvec_k=\zhat\cdot\mathrm{Ad}(G)\zhat$. Since the \mbox{$\zhat$-component} of $\mathrm{Ad}(G)\zhat$ equals \mbox{$|G_{11}|^2-|G_{21}|^2=1-2|G_{21}|^2$} by the $\SU$ structure, the per-mode residual collapses to a single off-diagonal entry,
\begin{equation}\label{eq:keystone}
  \varepsilon_k = 1-\bhat_k\cdot\mvec_k = 2\,|G_{21}(k)|^2 .
\end{equation}
Thus $\varepsilon_k=0$ if and only if $G(k)$ is diagonal: the steering condition \eqref{eq:gap} therefore becomes the demand that $G_{21}$ vanish at the $p$ physical momentum modes.

\subsection{\texorpdfstring{$G$}{G} as a QSP sequence}\label{sec:qsp}

Since the cost axis is a frame rotation of $\zhat$, each cost layer can be written as
\begin{equation}
  e^{-2i \gamma_{\ell}\, \bhat_k \cdot \tauv_k}
  =
  W_k^{\dagger} e^{-2i \gamma_{\ell} \tau_k^z} W_k.
\end{equation}
via $\bhat_k\cdot \tauv_k = W_k^{\dagger}\tau^z_k W_k$.
Using that $\tau^z_k W_k\tau^z_k=W_k^\dagger$ one can then bring the lifted propagator to the form
\begin{equation}\label{eq:qsp}
  G = (-1)^p\,W_k\prod_{j=2p}^{1}\big(e^{i\psi_j\tau^z_k}W_k\big),
\end{equation}
with
\begin{equation}
  \begin{aligned}
    \psi_{2\ell-1}&=-(2\gamma_\ell+\tfrac\pi2),\\
    \psi_{2\ell}&=-(2\beta_\ell+\tfrac\pi2),
  \end{aligned}
\end{equation}
which is identified as a standard QSP sequence \cite{Low:2016sck,Gilyen:2018khw}~\cite{LowYoderChuang2016} - an alternating product of the \emph{signal} $W_k$ and \mbox{$k$-independent} diagonal \emph{phases}.

Now define the signal variable $z:=e^{ik/2}$. In the eigenbasis of $\tau^y_k$ the frame rotation is linear in $z^{\pm1}$
\begin{equation}\label{eq:Wpoly}
  W_k = z\,E_+ + z^{-1}E_-, \qquad E_\pm=\tfrac12(I\mp\tau^y_k),
\end{equation}
with $E_\pm$ the spectral projectors of $\tau^y_k$.
$G$ is therefore an $\SU$-valued Laurent polynomial of degree $L:=2p+1$.
We can parametrize it as
\begin{equation}
    G = \Big(\begin{matrix}A&-B^*\\B&A^*\end{matrix}\Big).
\end{equation}
Unitarity imposes $|A|^2+|B|^2=1$ on the unit circle, and parity $G(1/z)=\tau^z_k G(z)\tau^z_k$ makes the diagonal even and the off-diagonal odd under the antipodal map $z\to1/z$. Defining $w:=z^2$, $A=z^{-L}R(w)$ and $B=z^{-L}T(w)$, we then have
\begin{equation}\label{eq:classL}
\begin{gathered}
  |R(w)|^2+|T(w)|^2=1,\\
  w^{L}R(1/w)=R(w),\qquad w^{L}T(1/w)=-T(w),
\end{gathered}
\end{equation}
with $R$ \emph{palindromic} and $T$ \emph{antipalindromic}.
We call a pair $(R,T)$ carrying this parity a
\emph{class-$L$} pair.

\subsection{The steering condition}\label{sec:achiev-cond}

By the keystone \eqref{eq:keystone}, the steering condition \eqref{eq:gap} becomes a clean statement about a single polynomial,
\begin{equation}\label{eq:achiev-poly}
  \varepsilon_k=0\ \forall k \iff T\ \text{vanishes at $p$ nodes}\ w_\ell=e^{ik_\ell},
\end{equation}
reducing simultaneous steering to a polynomial interpolation: optimality holds if and only if such a polynomial can be constructed.

\subsection{Determining the polynomial}\label{sec:forced}

As a class-$L$ entry, $T$ has a rigid analytic structure - a polynomial of degree $\le L$ in $w$, its only singularity a pole of order $L$ at $w=\infty$ - so it is fixed by its zeros up to a constant.

\paragraph{The zeros fix the form.}
The steering condition is now $T(w_\ell)=0$ at the $p$ nodes, and the class symmetry supplies the rest. Antipalindromy $w^{L}T(1/w)=-T(w)$ forces a root at $w=1$ (since $T(1)=-T(1)$) and drags each node's mirror $\bar w_\ell = w_{M-\ell}$ as a root, so $\{w_\ell\}\cup\{\bar w_\ell\}\cup\{1\}$ is exactly the $M:=2p+2$ roots of unity except $-1$ - the mode $k=\pi$, whose cost and mixer axes are antiparallel and which is therefore unsteerable. Up to scale we therefore have
\begin{equation}\label{eq:Tforced}
  T(w) = c\,(w^{L}-w^{L-1}+\cdots+w-1) = c\,\frac{w^{M}-1}{w+1}.
\end{equation}

\paragraph{Unitarity fixes the scale.}
The scale is set at the unsteerable point: palindromy with $L$ odd forces $R(-1)=0$, so
$|R(-1)|^2+|T(-1)|^2=1$ gives $|T(-1)|=1$, and since $T(-1)=-cM$,
\begin{equation}\label{eq:cval}
  c=\frac1M=\frac{1}{2p+2}.
\end{equation}
The FGG value $1/(2p+2)$ thus emerges as the normalization of the unique node polynomial, fixed by unitarity at the one mode that cannot be steered - a structural explanation of a constant previously seen only in numerics. As a consistency check, $|T|\le1$ on the unit circle, since $|\sin(\tfrac M2\theta)|\le M|\cos(\theta/2)|$ for $M$ even, with equality only at $\theta=\pi$.

\subsection{The complementary polynomial}\label{sec:fejer}

To complete $T$ to a class-$L$ pair we need a palindromic partner $R$ with $|R|^2+|T|^2=1$ on the unit circle. Since $|R(w)|^2=1-|T|^2\ge0$, the \emph{Fej\'er--Riesz theorem} \cite{FejerRiesz,Riesz1916} guarantees a spectral square root $R$, palindromic because $1-|T|^2$ is invariant under $w\mapsto1/w$, so the complementary polynomial exists and $(R,T)$ is a certified class-$L$ $\SU$ pair. The freedom in this square root - the $p$ antipodal pairs - yields a $2^p$ multiplicity of admissible $R$, hence $2^p$ optimal angle sets, a degeneracy that is a consequence of the dynamical symmetry and was observed numerically in \cite{Mbeng:2019rat}.

\subsection{The inverse map}\label{sec:haah}

To recover angles one maps $(R,T)$ back to a QSP sequence.

\paragraph{Haah's factorization.}
Haah's theorem~\cite{Haah2019product} factors any $\SU$-valued Laurent polynomial of degree $L$ uniquely and constructively,
\begin{equation}\label{eq:haah}
\begin{gathered}
  G = E_0\prod_{j=1}^{L}(zP_j+z^{-1}Q_j),\\
  Q_j=I-P_j,\quad P_j\ \text{rank-one},\quad E_0\in\SU,
\end{gathered}
\end{equation}
an explicit degree-lowering peel, so the primitives can be computed.

\paragraph{Parity fixes the projectors.}
Our extra parity $G(1/z)=\tau^z_k G(z)\tau^z_k$ forces $Q_j=\tau^z_k P_j\tau^z_k$, whose only solutions are \emph{equatorial},
\begin{equation}\label{eq:equatorial}
  P_j=\tfrac12(I+\cos\varphi_j\,\tau^x_k+\sin\varphi_j\,\tau^y_k),
\end{equation}
a one-parameter family, one phase $\varphi_j$ per primitive, with $E_0$ diagonal.

\paragraph{From primitives to angles.}
Each equatorial primitive is the fixed signal $W_k$ dressed by a diagonal phase $D(t):=e^{it\tau^z_k}$,
\begin{equation}\label{eq:letter}
  zP_\varphi+z^{-1}Q_\varphi = D(t_\varphi)\,W_k\,D(-t_\varphi),
  \qquad t_\varphi=\tfrac{\varphi}{2}+\tfrac{\pi}{4},
\end{equation}
so the product \eqref{eq:haah} takes exactly the shape of the QSP circuit \eqref{eq:qsp}, with internal phases $\psi_j=t_{\varphi_{j+1}}-t_{\varphi_j}=\tfrac12(\varphi_{j+1}-\varphi_j)$. Composing with the affine relations of~\eqref{eq:qsp} gives the dictionary to the QAOA angles,
\begin{align}\label{eq:angles}
  \gamma_\ell&=-\tfrac14(\varphi_{2\ell}-\varphi_{2\ell-1})-\tfrac{\pi}{4},\nonumber\\
  \beta_\ell&=-\tfrac14(\varphi_{2\ell+1}-\varphi_{2\ell})-\tfrac{\pi}{4},
  \qquad \ell=1,\dots,p.
\end{align}
The map is affine and explicitly invertible, completing the proof that optimal angles exist.

\section{Conclusion}\label{sec:conclusion}

We have completed the proof of the FGG conjecture for the ring of disagrees: for $n$ even and $2p+2\le n$, depth-$p$ QAOA attains an approximation ratio $(2p+1)/(2p+2)$ exactly. The proof is machine-verified in Lean~4.

The proof the model produced has a clear physical structure. It reduces the problem through the free-fermion symmetry to one $\SU$ system per mode and recognizes that system as a QSP sequence. This route is new to the problem: none of the papers on the conjecture \cite{Farhi:2014ych,Wang:2018rtf,Mbeng:2019rou,Mbeng:2019rat,Bravyi:2019jtd} use QSP, the QSP ``grand unification'' \cite{Martyn:2021eaf} does not list QAOA, and the nearest result, that transverse-field Ising \emph{dynamics} are QSP sequences \cite{PhysRevB.109.014306}, was not aimed at QAOA optimization. The construction is not special to this model: the keystone~\eqref{eq:keystone} follows from the per-mode $\mathrm{SU}(2)$ structure alone, so any translation-invariant free-fermion~\cite{Chapman2020graph,Chapman2023unified} QAOA whose per-mode dynamics close into $\mathfrak{su}(2)$ reduces to an analogous single-polynomial QSP steering problem.

We believe the method extends well beyond this problem. Within quantum information science, natural targets include conjectures in quantum computation~\cite{Preskill2018nisq}, the security analysis of quantum communication~\cite{Pirandola2020advances}, and secure computation~\cite{Sulimany2025qsmdl}. More broadly, any well-posed problem across mathematics, the physical sciences, computer science, and engineering, once formalized, can be tackled in the same fashion using an interactive theorem prover such as Lean.

Formal problems are a natural first domain for such autonomous discovery precisely because they are \emph{closed}: the kernel is a complete, deterministic, in-loop oracle, independent of the model that generates the argument, so the knowledge the model must get right reduces to the finite axioms and definitions fixed in the statement - a bounded ``world model'' a human audits once, within which a hallucinated step cannot survive compilation and is discarded rather than propagated.

\section*{Acknowledgments}
We would like to thank Edward Farhi for insightful conversations that motivated this work. We thank Kunal Marwaha for his comments on the draft and for a conversation on our independently derived results.

K.S.\ acknowledges support from the European Union's Horizon Europe research and innovation programme under the Marie Sk\l{}odowska-Curie grant agreement No.~101202109. The research of M.B.S.\ is supported by the Knut and Alice Wallenberg Foundation, Grant KAW 2023.0490.

\textit{Note added.} - During the preparation of the manuscript we became aware of an independent work on the same optimization problem~\cite{Marwaha_toappear}.

\section*{Code and data availability}
The complete Lean library, the formalized known components, and Fable~5's proof are available in the repository~\cite{repo}.

\bibliographystyle{apsrev4-2}
\bibliography{FGGring}

@misc{Farhi:2014ych,
    author = "Farhi, Edward and Goldstone, Jeffrey and Gutmann, Sam",
    title = "{A Quantum Approximate Optimization Algorithm}",
    eprint = "1411.4028",
    archivePrefix = "arXiv",
    primaryClass = "quant-ph",
    reportNumber = "MIT-CTP/4610, MIT-CTP/4610",
    month = "11",
    year = "2014"
}

@misc{repo,
      author       = "Kol, Uri and Ben-Shahar, Maor and Sulimany, Kfir and Englund, Dirk",
      title        = "{QuantumOptimization: a Lean~4 / Mathlib library of
                      machine-verified quantum-optimization results}",
      howpublished = "\url{https://github.com/urikol/QuantumOptimization}",
      year         = "2026"
  }

@unpublished{Marwaha_toappear,
      author = "Marwaha, Kunal",
      title  = "{Independent work on the ring of disagrees}",
      note   = "to appear",
      year   = "2026"
  }

@inproceedings{deMouraUllrich2021,
  title     = {The {Lean 4} Theorem Prover and Programming Language},
  author    = {de Moura, Leonardo and Ullrich, Sebastian},
  booktitle = {International Conference on Automated Deduction},
  pages     = {625--643},
  year      = {2021},
  organization = {Springer}
}

@inproceedings{mathlib2020,
  author = {{The mathlib Community}},
  title = {The {L}ean {M}athematical {L}ibrary},
  booktitle = {Proceedings of the 9th {ACM} {SIGPLAN} International Conference on Certified Programs and Proofs},
  year = {2020},
  doi = {10.1145/3372885.3373824}
}

@article{Wang:2018rtf,
    author = "Wang, Zhihui and Hadfield, Stuart and Jiang, Zhang and Rieffel, Eleanor G.",
    title = "{Quantum approximate optimization algorithm for MaxCut: A fermionic view}",
    eprint = "1706.02998",
    archivePrefix = "arXiv",
    primaryClass = "quant-ph",
    doi = "10.1103/PhysRevA.97.022304",
    journal = "Phys. Rev. A",
    volume = "97",
    number = "2",
    pages = "022304",
    year = "2018"
}

@misc{Mbeng:2019rou,
    author = "Mbeng, Glen Bigan and Fazio, Rosario and Santoro, Giuseppe",
    title = "{Quantum Annealing: a journey through Digitalization, Control, and hybrid Quantum Variational schemes}",
    eprint = "1906.08948",
    archivePrefix = "arXiv",
    primaryClass = "quant-ph",
    month = "6",
    year = "2019"
}

@misc{Mbeng:2019rat,
    author = "Mbeng, Glen Bigan and Fazio, Rosario and Santoro, Giuseppe E.",
    title = "{Optimal quantum control with digitized Quantum Annealing}",
    eprint = "1911.12259",
    archivePrefix = "arXiv",
    primaryClass = "quant-ph",
    month = "11",
    year = "2019"
}

@article{Bravyi:2019jtd,
    author = "Bravyi, Sergey and Kliesch, Alexander and Koenig, Robert and Tang, Eugene",
    title = "{Obstacles to Variational Quantum Optimization from Symmetry Protection}",
    eprint = "1910.08980",
    archivePrefix = "arXiv",
    primaryClass = "quant-ph",
    doi = "10.1103/PhysRevLett.125.260505",
    journal = "Phys. Rev. Lett.",
    volume = "125",
    number = "26",
    pages = "260505",
    year = "2020"
}

@article{Low:2016sck,
    author = "Low, Guang Hao and Chuang, Isaac L.",
    title = "{Optimal Hamiltonian Simulation by Quantum Signal Processing}",
    eprint = "1606.02685",
    archivePrefix = "arXiv",
    primaryClass = "quant-ph",
    doi = "10.1103/PhysRevLett.118.010501",
    journal = "Phys. Rev. Lett.",
    volume = "118",
    number = "1",
    pages = "010501",
    year = "2017"
}

@inproceedings{Gilyen:2018khw,
    author = "Gily{\'e}n, Andr{\'a}s and Su, Yuan and Low, Guang Hao and Wiebe, Nathan",
    title = "{Quantum singular value transformation and beyond: exponential improvements for quantum matrix arithmetics}",
    booktitle = "{51st Annual ACM SIGACT Symposium on Theory of Computing}",
    eprint = "1806.01838",
    archivePrefix = "arXiv",
    primaryClass = "quant-ph",
    doi = "10.1145/3313276.3316366",
    month = "6",
    year = "2018"
}

@article{Haah2019product,
  doi = {10.22331/q-2019-10-07-190},
  url = {https://doi.org/10.22331/q-2019-10-07-190},
  title = {Product {D}ecomposition of {P}eriodic {F}unctions in {Q}uantum {S}ignal {P}rocessing},
  author = {Haah, Jeongwan},
  journal = {{Quantum}},
  issn = {2521-327X},
  publisher = {{Verein zur F{\"{o}}rderung des Open Access Publizierens in den Quantenwissenschaften}},
  volume = {3},
  pages = {190},
  month = oct,
  year = {2019}
}

@article{PhysRevB.109.014306,
  title = {Quantum signal processing with the one-dimensional quantum Ising model},
  author = {Bastidas, V. M. and Zeytino\ifmmode \breve{g}\else \u{g}\fi{}lu, S. and Rossi, Z. M. and Chuang, I. L. and Munro, W. J.},
  journal = {Phys. Rev. B},
  volume = {109},
  issue = {1},
  pages = {014306},
  numpages = {15},
  year = {2024},
  month = {Jan},
  publisher = {American Physical Society},
  doi = {10.1103/PhysRevB.109.014306},
  url = {https://link.aps.org/doi/10.1103/PhysRevB.109.014306}
}

@article{Martyn:2021eaf,
    author = "Martyn, John M. and Rossi, Zane M. and Tan, Andrew K. and Chuang, Isaac L.",
    title = "{Grand Unification of Quantum Algorithms}",
    eprint = "2105.02859",
    archivePrefix = "arXiv",
    primaryClass = "quant-ph",
    doi = "10.1103/PRXQuantum.2.040203",
    journal = "PRX Quantum",
    volume = "2",
    number = "4",
    pages = "040203",
    year = "2021"
}

@article{Larocca:2021ksf,
    author = "Larocca, Martin and Czarnik, Piotr and Sharma, Kunal and Muraleedharan, Gopikrishnan and Coles, Patrick J. and Cerezo, M.",
    title = "{Diagnosing Barren Plateaus with Tools from Quantum Optimal Control}",
    eprint = "2105.14377",
    archivePrefix = "arXiv",
    primaryClass = "quant-ph",
    reportNumber = "LA-UR-21-24973",
    doi = "10.22331/q-2022-09-29-824",
    journal = "Quantum",
    volume = "6",
    pages = "824",
    year = "2022"
}

@article{Matos:2022koi,
    author = "Matos, Gabriel and Self, Chris N. and Papi{\'c}, Zlatko and Meichanetzidis, Konstantinos and Dreyer, Henrik",
    title = "{Characterization of variational quantum algorithms using free fermions}",
    eprint = "2206.06400",
    archivePrefix = "arXiv",
    primaryClass = "quant-ph",
    doi = "10.22331/q-2023-03-30-966",
    journal = "Quantum",
    volume = "7",
    pages = "966",
    year = "2023"
}

@article{FejerRiesz,
      author  = "Fej{\'e}r, Leopold",
      title   = "{\"U}ber trigonometrische Polynome",
      journal = "J. Reine Angew. Math.",
      volume  = "146",
      pages   = "53--82",
      year    = "1916",
      doi     = "10.1515/crll.1916.146.53"
  }

@article{Riesz1916,
      author  = "Riesz, Frigyes",
      title   = "{\"U}ber ein Problem des Herrn Carath{\'e}odory",
      journal = "J. Reine Angew. Math.",
      volume  = "146",
      pages   = "83--87",
      year    = "1916",
      doi     = "10.1515/crll.1916.146.83"
  }

@misc{leanscriber,
      title        = "{LeanScriber: an agentic toolkit for Lean~4 formalization}",
      howpublished = "\url{https://github.com/urikol/leanscriber}",
      year         = "2026"
  }

@article{AlphaProof2025,
      author  = "Hubert, Thomas and others",
      title   = "{Olympiad-level formal mathematical reasoning with reinforcement learning}",
      journal = "Nature",
      volume  = "651",
      pages   = "607--613",
      year    = "2025",
      doi     = "10.1038/s41586-025-09833-y"
  }

@article{Trinh2024alphageometry,
      author  = "Trinh, Trieu H. and Wu, Yuhuai and Le, Quoc V. and He, He and Luong, Thang",
      title   = "{Solving olympiad geometry without human demonstrations}",
      journal = "Nature",
      volume  = "625",
      pages   = "476--482",
      year    = "2024",
      doi     = "10.1038/s41586-023-06747-5"
  }

@article{RomeraParedes2024funsearch,
      author  = "Romera-Paredes, Bernardino and others",
      title   = "{Mathematical discoveries from program search with large language models}",
      journal = "Nature",
      volume  = "625",
      pages   = "468--475",
      year    = "2024",
      doi     = "10.1038/s41586-023-06924-6"
  }

@inproceedings{Wu2022autoformalization,
      author    = "Wu, Yuhuai and Jiang, Albert Q. and Li, Wenda and Rabe, Markus N. and Staats, Charles and Jamnik, Mateja and Szegedy, Christian",
      title     = "{Autoformalization with Large Language Models}",
      booktitle = "Advances in Neural Information Processing Systems (NeurIPS)",
      year      = "2022",
      eprint    = "2205.12615",
      archivePrefix = "arXiv"
  }

@inproceedings{Yang2023leandojo,
      author    = "Yang, Kaiyu and Swope, Aidan M. and Gu, Alex and Chalamala, Rahul and Song, Peiyang and Yu, Shixing and Godil, Saad and Prenger, Ryan and Anandkumar, Anima",
      title     = "{LeanDojo: Theorem Proving with Retrieval-Augmented Language Models}",
      booktitle = "Advances in Neural Information Processing Systems (NeurIPS), Datasets and Benchmarks Track",
      year      = "2023",
      eprint    = "2306.15626",
      archivePrefix = "arXiv"
  }

@misc{DeepSeekProverV2,
      author = "Ren, Z. Z. and others",
      title  = "{DeepSeek-Prover-V2: Advancing Formal Mathematical Reasoning via Reinforcement Learning for Subgoal Decomposition}",
      eprint = "2504.21801",
      archivePrefix = "arXiv",
      primaryClass = "cs.CL",
      year   = "2025"
  }

@article{Wurtz2021guarantees,
      author  = "Wurtz, Jonathan and Love, Peter",
      title   = "{MaxCut quantum approximate optimization algorithm performance guarantees for $p>1$}",
      journal = "Phys. Rev. A",
      volume  = "103",
      pages   = "042612",
      year    = "2021",
      doi     = "10.1103/PhysRevA.103.042612"
  }

@article{Wiersema2024dla,
      author  = "Wiersema, Roeland and K{\"o}kc{\"u}, Efekan and Kemper, Alexander F. and Bakalov, Bojko N.",
      title   = "{Classification of dynamical Lie algebras of 2-local spin systems on linear, circular and fully connected topologies}",
      journal = "npj Quantum Information",
      volume  = "10",
      pages   = "110",
      year    = "2024",
      doi     = "10.1038/s41534-024-00900-2"
  }

@article{Chapman2020graph,
      author  = "Chapman, Adrian and Flammia, Steven T.",
      title   = "{Characterization of solvable spin models via graph invariants}",
      journal = "Quantum",
      volume  = "4",
      pages   = "278",
      year    = "2020",
      doi     = "10.22331/q-2020-06-04-278"
  }

@misc{Chapman2023unified,
      author = "Chapman, Adrian and Elman, Samuel J. and Mann, Ryan L.",
      title  = "{A Unified Graph-Theoretic Framework for Free-Fermion Solvability}",
      eprint = "2305.15625",
      archivePrefix = "arXiv",
      year   = "2023"
  }

@article{Ebadi2022,
      author  = "Ebadi, S. and others",
      title   = "{Quantum optimization of maximum independent set using Rydberg atom arrays}",
      journal = "Science",
      volume  = "376",
      number  = "6598",
      pages   = "1209--1215",
      year    = "2022",
      doi     = "10.1126/science.abo6587"
  }

@misc{Pichler2018,
      author = "Pichler, Hannes and Wang, Sheng-Tao and Zhou, Leo and Choi, Soonwon and Lukin, Mikhail D.",
      title  = "{Quantum Optimization for Maximum Independent Set Using Rydberg Atom Arrays}",
      eprint = "1808.10816",
      archivePrefix = "arXiv",
      primaryClass = "quant-ph",
      year   = "2018"
  }

@article{Harrigan2021,
      author  = "Harrigan, Matthew P. and others",
      title   = "{Quantum approximate optimization of non-planar graph problems on a planar superconducting processor}",
      journal = "Nature Physics",
      volume  = "17",
      number  = "3",
      pages   = "332--336",
      year    = "2021",
      doi     = "10.1038/s41567-020-01105-y"
  }

@article{Bernien2017,
      author  = "Bernien, Hannes and others",
      title   = "{Probing many-body dynamics on a 51-atom quantum simulator}",
      journal = "Nature",
      volume  = "551",
      pages   = "579--584",
      year    = "2017",
      doi     = "10.1038/nature24622"
  }

@article{JordanWigner1928,
      author  = "Jordan, Pascual and Wigner, Eugene",
      title   = "{{\"U}ber das Paulische {\"A}quivalenzverbot}",
      journal = "Z. Phys.",
      volume  = "47",
      pages   = "631--651",
      year    = "1928",
      doi     = "10.1007/BF01331938"
  }

@article{Anderson1958,
      author  = "Anderson, P. W.",
      title   = "{Random-Phase Approximation in the Theory of Superconductivity}",
      journal = "Phys. Rev.",
      volume  = "112",
      pages   = "1900--1916",
      year    = "1958",
      doi     = "10.1103/PhysRev.112.1900"
  }

@misc{Farhi2020needs,
      author = "Farhi, Edward and Gamarnik, David and Gutmann, Sam",
      title  = "{The Quantum Approximate Optimization Algorithm Needs to See the Whole Graph: A Typical Case}",
      eprint = "2004.09002",
      archivePrefix = "arXiv",
      primaryClass = "quant-ph",
      year   = "2020"
  }

@article{LowYoderChuang2016,
      author  = "Low, Guang Hao and Yoder, Theodore J. and Chuang, Isaac L.",
      title   = "{Methodology of Resonant Equiangular Composite Quantum Gates}",
      journal = "Phys. Rev. X",
      volume  = "6",
      pages   = "041067",
      year    = "2016",
      doi     = "10.1103/PhysRevX.6.041067"
  }

@article{LiebSchultzMattis1961,
      author  = "Lieb, Elliott and Schultz, Theodore and Mattis, Daniel",
      title   = "{Two soluble models of an antiferromagnetic chain}",
      journal = "Ann. Phys.",
      volume  = "16",
      pages   = "407--466",
      year    = "1961",
      doi     = "10.1016/0003-4916(61)90115-4"
  }

@article{HoHsieh2019,
      author  = "Ho, Wen Wei and Hsieh, Timothy H.",
      title   = "{Efficient variational simulation of non-trivial quantum states}",
      journal = "SciPost Phys.",
      volume  = "6",
      pages   = "029",
      year    = "2019",
      doi     = "10.21468/SciPostPhys.6.3.029"
  }

@inproceedings{DraftSketchProve,
      author    = "Jiang, Albert Q. and Welleck, Sean and Zhou, Jin Peng and Li, Wenda and Liu, Jiacheng and Jamnik, Mateja and Lacroix, Timoth{\'e}e and Wu, Yuhuai and Lample, Guillaume",
      title     = "{Draft, Sketch, and Prove: Guiding Formal Theorem Provers with Informal Proofs}",
      booktitle = "International Conference on Learning Representations (ICLR)",
      year      = "2023",
      eprint    = "2210.12283",
      archivePrefix = "arXiv"
  }

@misc{AxProver2025,
      author = "Breen, Benjamin and Del Tredici, Marco and McCarran, Jacob and Aspuru Mijares, Javier and Yin, Weichen Winston and Sulimany, Kfir and Taylor, Jacob M. and Koppens, Frank H. L. and Englund, Dirk",
      title  = "{Ax-Prover: A Deep Reasoning Agentic Framework for Theorem Proving in Mathematics and Quantum Physics}",
      eprint = "2510.12787",
      archivePrefix = "arXiv",
      primaryClass = "cs.AI",
      year   = "2025"
  }

@article{Keesling2019,
      author  = "Keesling, A. and others",
      title   = "{Quantum Kibble-Zurek mechanism and critical dynamics on a programmable Rydberg simulator}",
      journal = "Nature",
      volume  = "568",
      pages   = "207--211",
      year    = "2019",
      doi     = "10.1038/s41586-019-1070-1"
  }

@article{Nguyen2023,
      author  = "Nguyen, Minh-Thi and others",
      title   = "{Quantum Optimization with Arbitrary Connectivity Using Rydberg Atom Arrays}",
      journal = "PRX Quantum",
      volume  = "4",
      pages   = "010316",
      year    = "2023",
      doi     = "10.1103/PRXQuantum.4.010316"
  }

@unpublished{benshahar_qcrypto,
  author = {Ben Shahar, Maor and Morag, Jonathan and Englund, Dirk and Sulimany, Kfir},
  title  = {A Machine-Verified Proof of Quantum Cryptography},
  note   = {In preparation},
  year   = {2026}
}

@article{Preskill2018nisq,
  title   = {Quantum Computing in the {NISQ} era and beyond},
  author  = {Preskill, John},
  journal = {Quantum}, volume = {2}, pages = {79}, year = {2018},
  doi     = {10.22331/q-2018-08-06-79}
}

@article{Pirandola2020advances,
  title   = {Advances in quantum cryptography},
  author  = {Pirandola, S. and Andersen, U. L. and Banchi, L. and Berta, M. and Bunandar, D. and Colbeck, R. and Englund, D. and Gehring, T. and Lupo, C. and Ottaviani, C. and Pereira, J. and Razavi, M. and Shaari, J. S. and Tomamichel, M. and Usenko, V. C. and Vallone, G. and Villoresi, P. and Wallden, P.},
  journal = {Adv. Opt. Photon.}, volume = {12}, pages = {1012--1236}, year = {2020},
  doi     = {10.1364/AOP.361502}
}

@article{Sulimany2025qsmdl,
  title   = {Quantum-secure multiparty deep learning},
  author  = {Sulimany, Kfir and Vadlamani, Sri Krishna and Hamerly, Ryan and Iyengar, Prahlad and Englund, Dirk},
  journal = {Phys. Rev. X}, volume = {15}, pages = {041056}, year = {2025},
  doi     = {10.1103/k8wg-qmbh}
}

\end{document}